# Data science in public health: building next generation capacity

Authors: N. Mirin[1], H. Mattie[2], L. Jackson[3], Z. Samad[4], R. Chunara[5,6*]

[1]Department of Social and Behavioral Sciences, School of Global Public Health, New York University, New York USA
[2]Department of Biostatistics, Harvard T.H. Chan School of Public Health, Harvard University, Boston USA
[3]Department of Pediatrics and Child Health, College of Medicine, Howard University, Washington D.C. USA
[4]Department of Medicine, Aga Khan University, Karachi, Pakistan
[5]Department of Biostatistics, School of Global Public Health, New York University, New York USA
[6]Department of Computer Science and Engineering, Tandon School of Engineering, New York University, Brooklyn, USA
*corresponding author

## Abstract

Rapidly evolving technology, data and analytic landscapes are permeating many fields and professions. In public health, the need for data science skills including data literacy is particularly prominent given both the potential of novel data types and analysis methods to fill gaps in existing public health research and intervention practices, as well as the potential of such data or methods to perpetuate or augment health disparities. Through a review of public health courses and programs at the top 10 U.S. and globally ranked schools of public health, this article summarizes existing educational efforts in public health data science. These existing practices serve to inform efforts for broadening such curricula to further schools and populations. Data science ethics course offerings are also examined in context of assessing how population health principles can be blended into training across levels of data involvement to augment the traditional core of public health curricula. Parallel findings from domestic and international 'outside the classroom' training programs are also synthesized to advance approaches for increasing diversity in public health data science. Based on these program reviews and their synthesis, a four-point formula is distilled for furthering public health data science education efforts, toward development of a critical and inclusive mass of practitioners with fluency to leverage data to advance goals of public health and improve quality of life in the digital age.

## 1. Data and Public Health

Globally, data is becoming a plentiful resource. Researchers and practitioners in public and population health are seizing opportunities to use these increased data resources alongside advances in machine learning and artificial intelligence (AI) to address problems – such as predicting infectious disease spread using mobility measured through mobile phone data (Wesolowski et al., 2015), measuring environmental exposures such as air pollution via satellite imagery and deep learning (Weichenthal et al., 2020), and understanding sub-daily lifestyle attributes and behaviors through passively mined social media data (Liu et al., 2017). However, algorithmic advances via machine learning and AI approaches are insufficient tools when deployed without a complementary theoretical and ethical framework, a paradigm with potentially dangerous implications for the deployment of data-driven solutions and care in public health (Richardson, 2020). The availability and fidelity of data sourced in the field is likewise not without challenges and imperfections that need to be addressed. Accordingly, we as practitioners need to grow capacity in relevant interdisciplinary expertise for using data while also addressing these challenges in close synergy with public health priorities, such as equity and socio-ecological perspectives. The key findings and recommendations of the National Academies of Sciences report on *Envisioning the data science discipline*, include a call to equip students with "data acumen" (National Academies of Sciences, Engineering, and Medicine, 2018). This data acumen entails getting meaningful, correct, and useful answers from data. The report further indicates that this data fluency requires skills that are typically not fully developed in traditional mathematics, statistics, and computer science courses. Indeed, "data acumen" foregrounds the need for new data science skills, with particular attention to "data literacy" – defined as the skill of comprehending and critically evaluating real-world data across personal, academic, and professional contexts (Fontichiaro and Oehrli, 2016) – to be developed directly within the context of public health

In addition to realizing the urgent need for trainees in the age of digital medicine to develop data acumen that is rooted in the priorities and principles of public health, it is important to remember that public health is an inherently interdisciplinary field with room still for new perspectives. Public health draws on diverse knowledge across the natural and social sciences, economics, urban planning, and subdisciplines of expertise in the use of data through epidemiology and biostatistics, yet its offerings do not traditionally encompass courses that specifically address novel data competencies. In a forward-looking manner, the interdisciplinary public health ethos would naturally grow to incorporate computational sciences, to best capture and understand factors related to health and health outcomes. Notably, data and algorithmic challenges in public health transcend challenges in healthcare settings where there are more standardized forms of data (e.g., insurance claims, electronic health records) and thus, amongst other differences, public health may require wider expertise specific to data acquisition and management (Chunara et al., 2013). Moreover, as a global discipline, the purview of public health encompasses rural, urban, developing, and developed contexts. Data training must be designed flexibly to suit this wide spectrum of use cases.

Training the next generation of the public health workforce in data science skills and literacy in complement to their core competencies will ensure the development of data science pipelines that leverage strengths of the public health perspective to balance algorithmic performance with complex population health challenges (e.g., need for data from multiple walks of life, cost of mis-action at scale on false positives/negatives, integration with existing workflows, prioritizing disparities and ethics). Infusing public health training with data and algorithmic competencies will bring analytic expertise close to the data generation and decision-making processes, promoting rapid, necessary and sustainable iteration between data collection, analysis and action (Chunara et al., 2017; Rehman et al., 2018). Those versed in public health priorities will be able to proactively address these challenges while fulfilling data science functions such as optimizing model performance. Indeed, recent work has surfaced flaws in the use of machine learning approaches in a siloed way, such as algorithms built without an understanding of whom or what the data represent (Buolamwini & Gebru, 2018), limiting internal and external validity of conclusions drawn from analyses (Lazer et al., 2014). Moreover, models with mis-specified objectives can exacerbate health disparities even when efforts to be sensitive to disparity are made (Obermeyer et al., 2019).

## 2.    Current Public Health Data Science Curricula

There has been a recent increase in new courses and topic coverage of skills for modeling with big and deep data (i.e., data large in number of instances and features, or very rich or complex in content, respectively (Szczuka & Ślęzak, 2013)). This instruction goes beyond traditional and foundational data analytic concepts in epidemiology and biostatistics in public health programs. Cultivated skills include topics such as data visualization and communication, data management and assessment, workflow and reproducibility, communication, and ethical considerations when working with data. As a review of such changes to the data education landscape, we investigate data science curricula across the top 10 schools of public health in the United States and the top 10 global universities for social science and public health as ranked by the US News and World

Report (US News & World Report, 2021a, 2021b). Taken together these overlapping lists produced a sample size of *n* = 15 university programs in public health, three of them international, and 13 of the 15 housed within a school or faculty of public or population health, with Stanford and Oxford being the exceptions. The order these universities appear in Table 1 corresponds to the average of their domestic (US) and international rankings as programs in social sciences and public health, as determined by US News and World Report. Findings from examination of the curricula at these highly ranked schools can be used to inform best practices and identify gaps for expansion of such curricula beyond these top schools towards benefitting a more diverse student audience.

Though sometimes open to undergraduates, we found that public health data science coursework was almost uniformly listed with graduate level department codes. Thus, we limited the scope of this review to post-graduate professional training and certification programs, Master-level degrees, Doctoral studies, and their associated courses. To identify courses within these programs for further consideration we examined department- and/or concentration-specific websites to verify core and elective course requirements for graduation. We then located all relevant classes via university online directories and course look-up tools. This process produced an initial sample of 203 courses for review. Next, we located and examined syllabi and course descriptions for all 203 classes via university online directories and course look-up tools. To thoroughly assess these courses, we retrieved all available texts describing each course, including summaries posted by registrars, departmental web listings, and, when available, the most recent versions of syllabi. Examining the course topics and instructional approaches in detail led us to exclude 39 entries because the courses were, upon review by two of the authors, either considered standard public health instruction in biostatistics or epidemiology or found to focus on data types from basic science, such as genomics or biochemistry. This narrowed our sample of public health data science courses to *n* = 164.

Next, we performed a close qualitative review of the course descriptions, taking note of specific conceptual keywords and analysis types associated with novel data sources and applications. This classification system was needed because the boundary between biostatistics and data science can be indefinite in some ways (e.g., regression techniques foundational to statistical curricula may also be included in courses on machine learning/data science fundamentals). The system was used to identify courses which 1) clearly merited inclusion as public health data science-related instruction, 2) clearly merited exclusion (biostatistics training), or 3) required more detailed evaluation to reach an inclusion/exclusion determination. If the title of a course matched biostatistics topics and, upon qualitative review, its description(s) clearly focused on content taught within the traditional biostatistics discipline, we excluded it from our data science course list. For instance, courses teaching Bayesian methods were classified as biostatistics and excluded unless the methods in question were applied to machine learning explicitly, or if Bayesian fundamentals were reviewed only briefly before proceeding to direct applications of data science and machine learning, Conversely, we deemed courses focused on concepts including, but not limited to, the management and development of analytic methods specific to novel data types consistent with a 'public health data science' designation. By this token, if a course presumed familiarity with basic Bayesian methods as an enrollment prerequisite such that class time focused on advanced computing techniques, rather than cultivating familiarity with theoretical statistics, it was categorized as data science and included for analysis.

Beyond course-level classification, we also cataloged numbers and types of public health data science programs available at the universities under consideration. Programs were included for review if the name of the degree or certificate in question included a term explicitly denoting a focus on working with sources of health data through methods such as machine learning, AI, data science, informatics, data analytics, or spatial analysis. Degrees focused on traditional public health disciplines such as biostatistics, epidemiology, and health policy/management were also marked for inclusion, but only if they stipulated one or more public health data science course as a degree requirement, made public health data science courses available as either inter- or intra-departmental electives, or offered one or more public health data science course exclusively to certain students in that program (e.g., for doctoral candidates only). Incorporating this latitude into our inclusion criteria enabled us to capture programs actively endeavoring to weave data science skills and data literacy into their core curricula, as we advocate they should. It is worth noting that none of the public health programs under consideration required students to complete a 'data agnostic' data science skills or data literacy course drawing on case study content from outside the biomedical or population health domain. Although at the course level we included cross-listed electives with a 'data agnostic' focus in order to give credence to institutions with robust interdepartmental offerings for data work, at the program-level we honed our focus such that we could assess the state of data science integration into core public health curricula.

## 2.1 Course-level Analysis

To analyze the topics covered in public health data science curricula, we first synthesized categories that have been outlined in several articles on the competencies needed in data science writ large (Hardoon, 2021; Rodolfa & Ghani, 2021). Based on this analysis and our detailed syllabi review process, each of the 164 qualifying data science courses was classified with one of six comprehensive domain labels. The Machine Learning/AI label applied to 58 courses, or 35% of the 164 on offer across the degrees and short programs considered. Data Management (39 courses, 24%) was the second most common course type, with topics spanning data collection, organization, handling and cleaning, etc.). We designated the remaining courses either as Advanced Data Analysis (22 courses, 13%), which included instruction in methods such as spatial or temporal statistics, Data Use/Ethics (20 courses, 12%) on best practices for data use in health contexts, including regulatory and ethical considerations, Novel Data Handling (15 courses, 9%), with a focus on identifying and using data sources such as mobile health or other data from health systems, or Data Communication (10 courses, 6%) coursework on topics including data/information visualization, working with stakeholders, etc.).

*Findings from thematic analysis from course-level analysis*

The most common type of data science course in public health degree and programs focused on Machine Learning/AI (35%), with most included schools offering four or more courses on this topic (Figure 1). This domain label encompasses courses on topics spanning machine learning, AI, natural language processing, classification, and tree-based algorithms, etc. On the other hand, most schools offer only zero or one course in either Data Communication or Data Use/Ethics (Figure 2). Given redoubled illumination of the link between social factors and significant health disparities during COVID-19, as well as increasing popular, media, and congressional attention to potential embedded biases in Artificial Intelligence methods, we also further scrutinized the types of data science ethics content offered in the 20 courses that fall under the Data Use/Ethics category.

Based on our detailed manual review, we first found that Data Use/Ethics courses listed in public health data science programs could be categorized by content into instruction focused on: 1. accountable data management (14 of 20 courses, 70%), 2. privacy/confidentiality (4 of 20, 20%), 3. data sourcing processes (4 of 20, 20%) and 4. equitable delivery of care (4 of 20, 20%).

Notably, outlining and articulating these subcategories of instruction helped illustrate a gap in the way courses currently offered in public health data ethics remain detached to varying degrees from context surrounding the social and political dimensions of data and its use (Braveman, 2006). This includes the socioecological context that shapes different populations' needs, vulnerabilities, and access to health promoting resources across political, environmental, community, interpersonal, and intrapersonal levels. Such topics are integral to public health and commonly relayed through courses in socio-behavioral sciences, international health, and interdisciplinary coursework that draws on content from various social science disciplines such as political science, sociology, psychology, medical anthropology, and others.

## 2.2 Program-level Analysis

In addition to a course-level assessment, to better assess how students can avail themselves of public health data science curricula and in what depth, we also catalogued the types of programs through which data science is offered within the 15 included schools. Most frequently, these were on-campus, in-person Master's level programs (29 of them across the schools described above). Likewise, in our sample there were 11 PhD programs in public health offering specializations in data science or presenting data science course offerings. Despite their orientation toward big data and novel computational techniques, these Master's and PhD-level data science tracks are largely housed in epidemiology, biostatistics, health informatics, or similarly named, well-established public health departments. A small number (5) of short, certificate-granting training programs for working professionals and executives were also represented in our sample, alongside several (3) online Master's programs with flexible timelines and course loads meant to encourage lifelong learners to gain qualifications in data science topics at their own pace.

*Findings from thematic analysis from program-level analysis*
While encouraging that such a variety of programs are being offered, findings regarding the types of programs offering data science coursework demonstrate an opportunity to increase the variety of educational approaches to data science literacy to a wider group of public health students and practitioners. For example, students in public health disciplines that traditionally entail a relatively lesser focus on data analytics – such as socio-behavioral sciences – may find cultivating data science skills rooted in public health values, theories, and causal frameworks increasingly beneficial in today's data-driven environments. Our review of socio-behavioral science curricula descriptions at both the course- and program-level revealed that – across the universities under review – students in the socio-behavioral concentration largely did not have access within their home departments to data-intensive coursework taught with companion emphasis on the broader political, legal, and cultural contexts within which all data work is embedded (Braveman, 2016). When given the opportunity, these students may leverage the increased availability of data toward their research and practice ends even though they are not in degree-granting programs focused on biostatistics or epidemiology.

Overall, findings from the implementation landscape of data science coursework and programs in public health show that establishing public health data science degree programs and tracks within existing Master's/PhD programs is feasible, and such review can be useful to inform best practices and to identify current gaps in practice and scholarship. This is important both for schools undergoing curriculum review, as well as for expansion of such curricula beyond the top schools and towards benefitting a more diverse student audience. Our synthesis also highlights directions for further data science course development, specifically with respect to data science curricula in public health that can help drive the use of data science in mitigating and reducing health disparities, a key priority of public health (Liu et al., 2017).

## 3. Diversity improving entry points to public health data science

Given the importance of diversity in public health, in addition to the analyses presented above, here we describe tangible examples of ways to bring a diverse group of learners into public health data science. Research has shown that data science teams are frequently narrow in their provenance, with limited representation when it comes to including anyone from areas of the world beyond highly industrialized, financialized, and densely-networked locales (Sutherland, 2016). While graduate-level degree programs in Health Data Science are gradually emerging especially in European and North American Universities, universities outside those locations lack Health Data Science programs (Beyene et al., 2021). The importance of this focus is further exemplified by the fact that one of the International Statistical Institute's four strategic priorities is building statistical capacity in developing countries. Within the United States, Black students are dramatically underrepresented in computing and biostatistics fields. Black students make up only 2.0% of enrolled students in PhD programs in computing fields (Kuhlman et al. 2020) and in 2019 only 2.6% of Biostatistics degrees were awarded to Black students (Hidalgo et al. 2022).

As such, we draw on two trainings that were developed as exemplars to address the clear need to augment and encourage health data science capacity in diverse and marginalized groups. We then distill common threads between the programs and consider ways to support their effective implementation and expansion going forward. First, training and education formats must be flexible. Leveraging available educational formats such as online modules or short workshops in addition to supplemental coursework in degree curricula can an effective method to reach students at different stages and across different settings. For example, a short course on Data Science and Machine Learning in Public Health offered at New York University (NYU) and at Aga Khan University (AKU) – a large academic hospital in Karachi, Pakistan – demonstrated health data science capacity development in diverse groups both in a North American and global south setting. These were modular sessions providing an explanatory overview and practical examples of what the core topics are within data science and how they can be applied in public health. Trainees and students from a wide range of disciplines participated: medical students, residents, biostatisticians, and epidemiologists working in a healthcare setting attended at AKU workshop, while at NYU Master's and PhD students in social-behavioral sciences and nursing, as well as biostatistics and data science (not from public health) all took part. Beyond traditional methods in biostatistics and epidemiology only one out of the 27 total students and professionals in the trainings had been exposed to topics from a Data Science curriculum before. Though a short workshop of this nature cannot provide sufficient depth to cultivate expertise, it still serves to expose participants to the field of data science and suggest possible links to their ongoing work (prior to the workshop only 4 of the 27 total participants reported a 'complete' understanding of data science). Following the 5 days of training, 16 out of 21 respondents across both settings reported they had identified tangible ways they might integrate data science topics into their current coursework or research, post-workshop.

When seeking to broaden the accessibility of data science opportunities to students in degree programs that do not normally accommodate registration for a course beyond program requirements, calibrating the duration of the coursework and time of year of offering are key. The NYU and AKU workshops both took place close to or during typical summer/break times when competing responsibilities were somewhat reduced. It should also be noted that a short data science exposure, while not able to go into complete depth on all data science topics, was still reported to be an informative venue for all participants, particularly with respect to deciding whether they should pursue a deeper course in the future to fully realize the benefit data science approaches may have to their existing work. Indeed, students self-reported that opportunities for the collection and analysis of non-standard data could feasibly have a direct impact in their subfields of public health (e.g., acquiring measures of behavior, diet, and other health-related factors beyond what is possible to garner through surveys, or using machine learning to study complex relationships between such variables). Accordingly, for individuals coming from domains without traditional focus on data-driven methods, this brief data science exposure can provide them the initial impetus needed to feel comfortable delving further into the field, either through their work or by pursuing more formal education. Thus, we contend that more opportunities for such content exposure, outside of full courses, would be beneficial to the field. Proliferation of shorter course structures would likewise provide flexibility and encourage life-long learning opportunities. Those who are 5, 10 and more years out from their degree programs and embedded in organizations accruing and using data every day could benefit in both efficiency and insight by learning new concepts and best practices to deploy in their work even if they did not pursue data science studies in their original training.

Another approach which has demonstrated effectiveness in catalyzing broadening diversity among the ranks of individuals equipped with data science acumen is the Broadening Participation in Data Mining workshop (BPDM), currently overseen by professors at Howard University and a recent PhD from Nova Southeastern University, two historically Black and minority-facing institutions in the United States (Jackson & Maestre, 2020). The participatory format of BPDM focuses on both broadening exposure to data intensive topics and technologies and enriching technical aptitude and exposure for underrepresented groups by fostering mentorship, guidance, and network connections across academic, industry and governmental spaces. This fusion approach has been recognized as critical for the increased recruitment, retention, and productivity of underrepresented groups in science, technology, engineering, and mathematics. While BPDM activities are described in detail elsewhere (Jackson & Maestre, 2020), briefly, they encompass mentorship, reflection on the past, and goal-focused career planning. Beyond rote training activities, these workshops serve as venues to proactively support minoritized and marginalized students in their data science efforts, bolstering their self-concept while providing exposure to new collaboration and placement opportunities. Findings from the workshop series also reflected that participants felt greater connectedness to their discipline upon developing a solid network of peer and mentor colleagues. Over seven years, multiple iterations of this approach have consistently demonstrated that inclusion of topics related to public health data science can also help researchers capture and create authentic, culturally contextualized work that enhances diversity and nuance of perspective in science, technology, engineering, and mathematics fields (Carlton University, 2021; Friend, 2013). Given the white, male, industrialized skew within data science and other computational fields (Kuhlman et al., 2020), initiatives in public health data science education should welcome such efforts to broaden representation and modes of thinking in the emerging arena of public health data science practice.

## 4. Recommendations Summary and Conclusion

Based on findings from our review and efforts reported here, we outline a 4-point formula for schools undergoing curriculum review and/or establishing new programs in Public Health Data Science (summarized in Table 2). These recommendations are relevant both for public health departments as well as data science tracks or initiatives creating public health-focused programming. First, on a course-level, we note that programs can leverage principles from socio-behavioral sciences and other core public health disciplines within data science coursework, and in so doing advance discussion on the social and political dimensions of data and its use, both of which have consequences for public health. This will bring to bear a set of foundational principles from public health, such those articulating the boundaries and functions of systemic racism (Bailey et al., 2017), during data-centric discussions. We likewise recommend leveraging public health theory to inform work on "AI for good" (Chunara & Ralph, 2021) with principles from the study of and action upon inequity; a core tenet of the public health discipline (Richardson, 2020).

Second, to ensure that all those interested in infusing public health efforts with modern data science approaches are sufficiently informed, we must ensure that all competencies needed to participate in data science are made available to students. At present, data from our review show that coursework is often slanted heavily towards AI/machine learning applications, which may omit essential knowledge on data preparation, communication and use from students' toolboxes. Public health as a field must take caution not to train its learners to act on data in a vacuum. We further note that students within and outside of traditionally analytic and quantitative domains of public health (i.e., biostatistics and epidemiology) can benefit as scientists from exposure to a broad swathe of data science curricula.

Our third recommendation takes into account the nature and depth of public health data science offerings at a program level. Given the profile of programs currently offering public health data science curricula at the top 10 U.S. and international schools, as well as experiences reported by students and professionals across an array of public health areas, we find there is a pressing need for continuing learning opportunities. The fields of data science and digital medicine are evolving rapidly and taking mutually constitutive turns toward a data-fied world, and thus current professionals and researchers trained to work in public health may significantly benefit from the opportunity to understand how data science may be integrated into their work. Several schools in the included list (Table 1) offer short format courses and continuing learning opportunities. However, none of the trainings are focused on data science specifically within public health nor have been tailored for the continuing education of a public health audience, as we advocate.

Finally, as the value of diversity is recognized as integral to efforts in public health practice, our fourth recommendation is that diversity must be considered paramount in public health education efforts as well. In this paper we give an examples of approaches that have broadened representation in data science efforts, and advocate for further research into paradigms that will improve the diversity of the current public health workforce, both domestically and around the world.

The widespread training of individuals in true data acumen – core data science skills complemented by an openness to multidisciplinary perspectives – stands to grow the conceptual depth and interventional capacity of research and practice in public health as a field. Perhaps more importantly, increasing the number of field ambassadors with dual fluency in both data communication and population health principles may organically encourage critically informed and data literate approaches to population health to proliferate across the practice environment, generating a network effect of emanating advocacy. With their data acumen, these knowledge agents will be equipped to advance goals of health promotion in public discourse and motivate investment in disease prevention and improving quality of life, while also building trust in data-intensive AI and machine learning models among lay audiences. These ambassador roles constitute a vital civic function that public health graduates are well-positioned to fulfill and should be encouraged to pursue during multiple phases of their education. Through this review of existing efforts, we seek to augment the development of a critical, inclusive, and diverse mass of public health data science practitioners.

**Table 1. Top 10 schools of public health in the United States and top 10 global universities for social science and public health as ranked by US News & World Report.**

| Ranking US (Global) | University | Schools/Faculties/Institutes where Public Health Data Science Courses Located | PHDS Degree Programs | PHDS Courses Offered |
|---|---|---|---|---|
| 1 (2) | Johns Hopkins University | Johns Hopkins Bloomberg School of Public Health | 4 | 5 |
| -- (3) | Stanford University | Stanford School of Medicine | 7 | 17 |
| 3 (1) | Harvard University | Harvard T.H. Chan School of Public Health | 1 | 17 |
| -- (4) | University of Oxford | University of Oxford Medical Sciences Division/ Oxford Big Data Institute / Li Ka Shing Centre for Health Information and Discovery | 3 | 11 |
| -- (5) | University College London (UCL) | UCL Institute of Health Informatics ( within the Faculty of Population Health Sciences, at the School of Life & Medical Sciences) | 10 | 7 |
| 4 (6) | University of Michigan (UM) | UM School of Public Health | 2 | 6 |
| -- (10) | University of Toronto | Dalla Lana School of Public Health | 2 | 6 |
| 2 (9) | University of North Carolina, Chapel Hill | Gillings School of Global Public Health | 2 | 5 |
| 4 (8) | Columbia University | Columbia Mailman School of Public Health | 1 | 7 |
| 7 (7) | University of Washington | UW School of Public Health | 2 | 7 |
| 8 (14) | University of California, Berkeley | Berkeley School of Public Health | 3 | 5 |
| 10 (15) | University of California, Los Angeles (UCLA) | UCLA Fielding School of Public Health | 3 | 5 |
| 4 (40) | Emory University | Rollins School of Public Health | 5 | 16 |
| 10 (36) | University of Minnesota Twin Cities (UMN) | UMN School of Public Health | 4 | 10 |
| 8 (44) | Boston University | Boston University School of Public Health | 1 | 11 |

**Table 2. Four-point formula to enhance public health data science opportunities for diverse groups beyond standard coursework implementations.**

| Public Health Data Science Curriculum Recommendations |
|---|
| 1. *Socio-behavioral science principles* should be integrated into data science courses to advance discussion on the social and political dimensions of data and its use. |
| 2. *Courses across the data science spectrum* (advanced data analysis, machine learning/AI, data communication, data management, novel data handling and data use/ethics) should be available to students within data science-specific degree programs, tracks as well as traditionally less data-intensive concentrations. |
| 3. *Life-long learning opportunities* are necessary to enable public health practitioners to keep informed and skilled in the latest data science best practices. |
| 4. *Informal learning opportunities and mentorship* can be used to advance interdisciplinary connections and diversity in the public health field. |

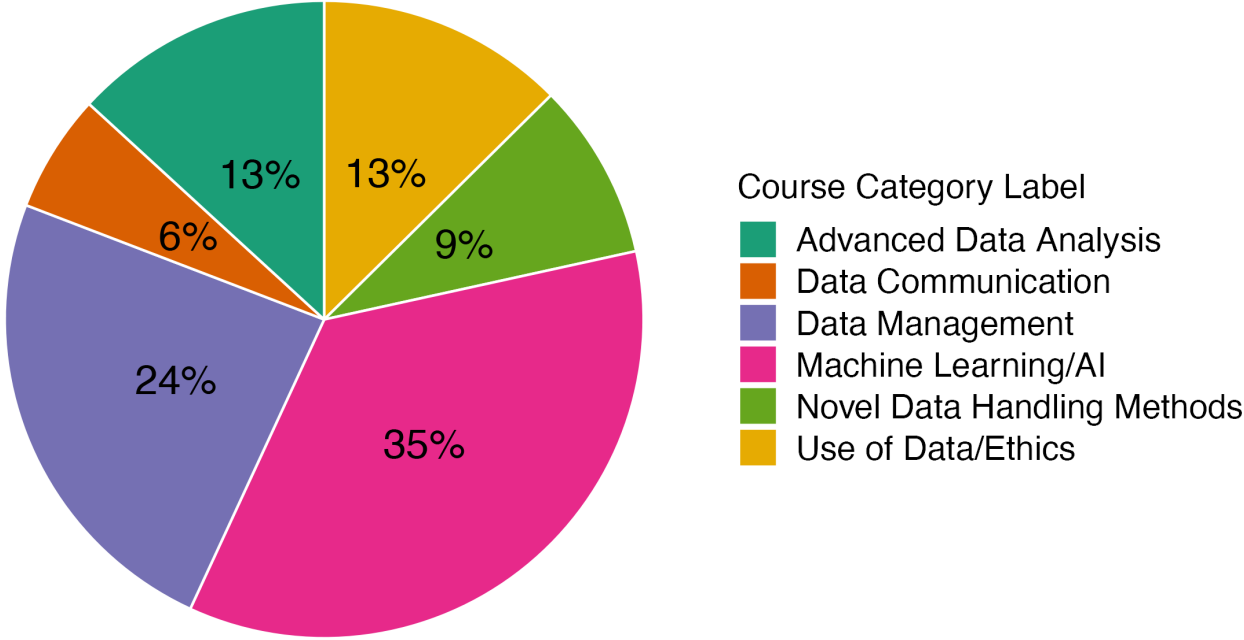

**Figure 1. Overall distribution of current coursework in public health data science programs.** Courses at the top 10 schools of public health in the United States and the top 10 global universities for social science and public health as ranked by the *US News and World Report* are categorized across six salient domains.

**Figure 2. Distribution of public health data science courses by category.**
Number of courses in each data science category (e.g., Machine learning/AI, etc.), in included schools.

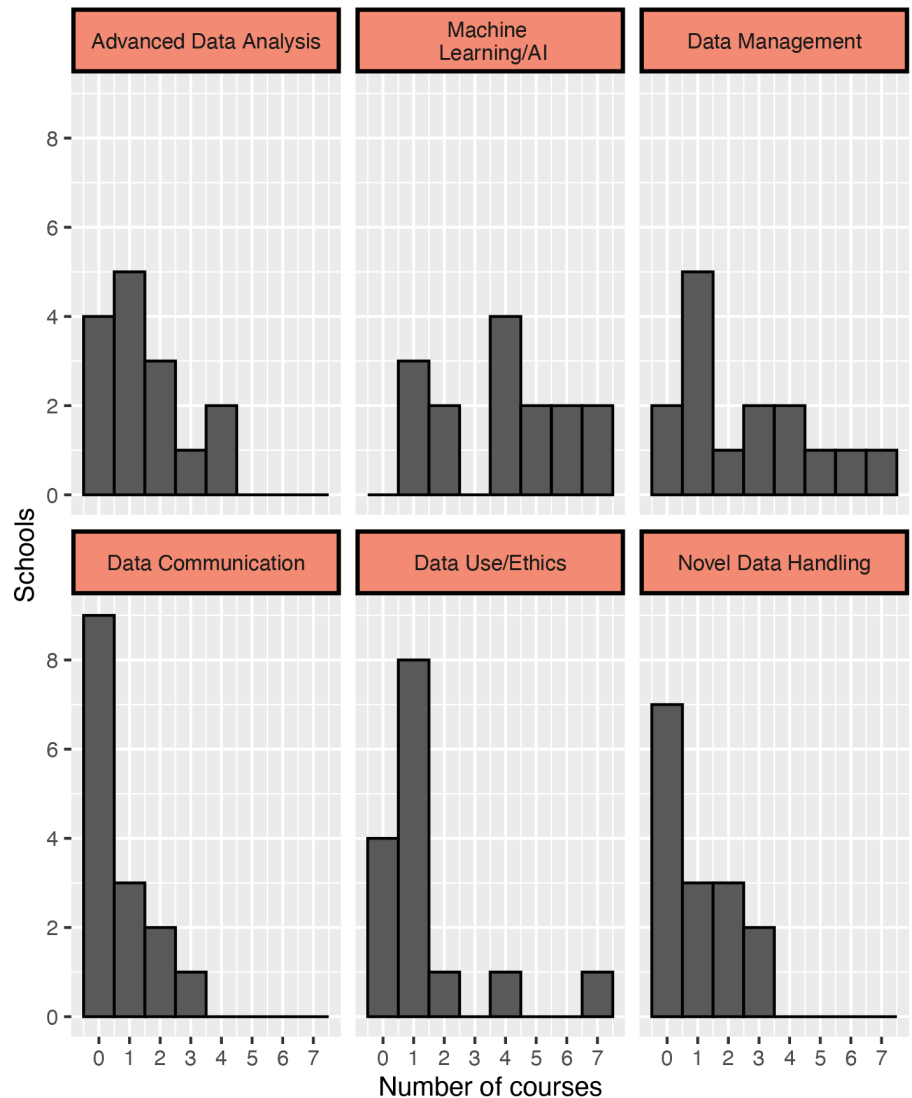